\documentstyle[12pt,titlepage]{article}

\font\grande=cmr10 scaled \magstep4
\font\medio=cmr10 scaled \magstep2
\outer\def\beginsection#1\par{\medbreak\bigskip
      \message{#1}\leftline{\bf#1}\nobreak\medskip
\vskip-\parskip
      \noindent}

\def\laq{\raise 0.4ex\hbox{$<$}\kern -0.8em\lower 0.62
ex\hbox{$\sim$}}

\def\gaq{\raise 0.4ex\hbox{$>$}\kern -0.7em\lower 0.62
ex\hbox{$\sim$}}
\begin{document}

\titlepage
\begin{flushright}
TUPT-03-99\\
april 1999\\
\end{flushright}
\vspace{2cm}
\begin{center}
{\grande Relic Gravitons, Dominant Energy Condition}\\ 
\vspace{4mm}
{\grande and Bulk Viscous Stresses  }\\

\vspace{20mm}
Massimo Giovannini
\footnote{Electronic address: giovan@cosmos2.phy.tufts.edu} \\
\vspace{6mm}

{\sl Institute of Cosmology, Department of Physics 
and Astronomy\\
Tufts University, Medford, Massachusetts 
02155, USA }

\end{center}

\vskip 2cm
\centerline{\medio  Abstract}

\noindent
If the 
 energy momentum tensor contains bulk viscous stresses violating the dominant
energy condition (DOC) 
the energy spectra of the relic gravitons (produced at the time 
of the DOC's violation)
increase in frequency in a calculable way. In a general relativistic context
we give examples where the DOC is only violated 
for a limited amount of time after which the ordinary (radiation dominated) 
evolution takes place. We connect our discussion to some recent 
remarks of Grishchuk concerning the detectability of the stochastic 
gravitational wave background by the forthcoming interferometric 
detectors.

\vspace{5mm}

\newpage

Every transition of the background geometry leads, inevitably, to the 
production of stochastically distributed relic gravitons whose energy spectra 
represent a crucial probe of the very early stages of the evolution of
our Universe \cite{1}. 
Prior to the  nucleosynthesis epoch there are no direct
 test of the thermodynamical state of the Universe and the presence of an
inflationary phase of expansion is usually justified  by causality arguments 
applied to the Cosmic Microwave Background (CMB) photons whose 
emission regions could not have been in causal contact in the far past 
if a never ending  radiation dominated phase would precede our present 
matter dominated stage of expansion \cite{2}. 

An inflationary evolution, if regarded at an effective level,  violates the 
(general relativistic) strong energy condition (SEC) namely 
$\rho + 3 p \geq 0$ where $\rho$ and $p$ are, respectively, the energy 
density and the pressure density of the perfect fluid sources 
driving the evolution of the background geometry \cite{3}. The need for 
such a violation can be immediately seen from the structure 
of Friedmann equations which imply that, if $\ddot{a} >0$, the SEC needs to be 
violated ($a(t)$ is the scale factor of the homogeneous and isotropic
Universe and the over-dot denotes the differentiation 
with respect to the cosmic time $t$).

Recently, Grishchuk \cite{4,5} 
 made an interesting observation concerning the 
slopes of the energy spectra of relic gravitons emerging 
from the models of the early Universe.  In short the argument goes 
as follows.  Suppose that we are in the framework of general 
relativity and suppose, as it is usually done, that the effective 
sources driving the background geometry can be parameterized by a 
stress tensor with perfect fluid form. As we stressed at the 
beginning, there are little doubts that the Universe had to 
be dominated by radiation (at least) since nucleosynthesis. Prior to that 
epoch we do not know which kind of energy momentum tensor could 
approximate the background sources but 
 we would like to deal with ever expanding Universes.
Moreover, at least for some time, we would like 
an accelerated expansion in order to solve the so called 
kinematical problems of the standard cosmological model. The second of the two 
previous requirements necessarily leads to the violation of the SEC.

A very naive model of the early Universe would then be given by 
two phases. An unknown phase where the perfect fluid sources have 
a generic equation of state $p= \gamma \rho$ , followed, at 
some transition time $t_1$, by a radiation dominated phase with 
$p = \rho/3$. The question we are very interested 
to ask is under which conditions 
such a toy model would produce energy spectra of relic gravitons
increasing faster than the first power of the frequency. Such a question 
is of obvious experimental relevance since, in the near future, various 
interferometric detectors of gravitational waves will come in operation. 
Now, if the energy spectra of relic gravitons are flat (or slowly increasing) 
with frequency there are little hopes of detecting them. In fact, the COBE 
limit applied to a frequency $\nu_0 \sim
  10^{-18}~h_0$ Hz imposes that 
the relic gravitons energy density (in critical units)
$h_0^2 \Omega_{{\rm GW}}$ has to be smaller 
then (or of the order of) $ 6.9 \times 10^{-11}$ ($h_0$ is the present 
indetermination in the value of the Hubble constant). Due to the 
transition from radiation to matter, the infra-red branch of the graviton 
spectrum declines (in frequency) as $\nu^{-2}$ between 
$\nu_0$ and $\nu_{2} \sim 10^{-16}~\Omega_0 ~h_{0}^2$ Hz.  
If we take into account that the typical frequency of operation of
 the interferometers
is between $10$ and $100$ Hz, we see that it is quite interesting to 
understand which kind of models could give  energy densities (around 
$100$ Hz)  larger than the typical inflationary models whose 
prediction is  $h_0^2 \Omega_{{\rm GW}} \leq 10^{-14}$. 

If we confine our attention to expanding Universes (i.e. $\dot{a} >0$) 
the slopes of the energy spectra are crucially related to the sign of 
$p + \rho $. This point can be easily seen by looking at the Friedmann 
equations 
whose solution gives $a(t)=t^{\alpha}$ with $\alpha = 2/[3( \gamma + 1)]$, 
having assumed $p = \gamma \rho$. The spectra of relic gravitons 
produced because of transition of the geometry from a generic perfect 
fluid stage to a radiation stage can be easily computed by matching the 
solutions of Eq. (\ref{grav}) in the two temporal regions defined by 
the transition time $t_1$. 
The requirement that we want expanding (and inflationary) Universes 
implies that the energy 
density of the created gravitons cannot increase with frequency if 
$p + \rho \geq 0$.

It was noticed by Barrow \cite{6} 
that a violation of the DOC does not necessarily 
forbid the viability of a given cosmological model in the same way as 
the violation of the SEC does not forbid the viability of ordinary inflationary
models \cite{7}. 
A violation of the DOC implies that $\dot{H}>0$
since, by combining the Friedmann equations we get that 
\begin{equation}
M^2_{P} \dot{H} = - \frac{3}{2} ( \rho + p).
\end{equation}
Notice, moreover, that the requirement of eternal expansion
implies, together with $\dot{H} >0$, that $\ddot{a} >0$.
In the context of general relativity, cosmological solutions 
violating the DOC (but not violating the weak energy conditions) have 
been sometimes connected with the presence of bulk viscous stresses 
driving the evolution of the geometry. 

Bulk viscous stresses \cite{8}
can appear quite naturally in cosmology as the result 
of the processes of particles
 production \cite{9} in curved space-time .
Moreover, bulk viscosity represents an interesting extension of the standard 
cosmological scenarios since it introduces dissipation in the model without 
breaking isotropy but by only redefining the space-space components 
of the energy momentum tensor according to 
\begin{equation}
p' = p - 3 \xi H,
\end{equation} 
where $\xi$ is the bulk viscosity coefficient. The Friedmann equations in 
the presence of bulk viscous stresses can be easily written as 
\begin{eqnarray}
&& M^2_{P} H^2 = \rho,
\nonumber\\
&& M^2_{P} ( \dot{H} + H^2) = - \frac{1}{2} ( \rho + 3 p'),
\nonumber\\
&& \dot{\rho} + 3 H ( \rho + p') =0.
\label{fr2}
\end{eqnarray}
As an example of the possibility of violating the DOC in not 
unreasonable models let us assume the simplest possibility namely 
$\xi = {\rm const} >0$ and $\gamma = -1$ which represents a particular case of 
the solution described by Barrow. The scale factor associated with this 
solution has a typical double exponential form \cite{6}
\begin{equation}
a(t) =a_0 \exp{[ \exp{( \frac{9}{2} \xi t)}]}
\label{sol1}
\end{equation}
and it describes an accelerated expansion with $\dot{H}>0$ (we took $M_P =1$). 
In the context of \cite{6} it was argued that this behavior may be 
connected with a violation of the second principle of thermodynamics 
generalized to curved backgrounds (of quasi-de Sitter and non de Sitter 
type \cite{6b}). 
Other solutions of the type given by Eq. (\ref{sol1}) 
can be constructed by assuming, for instance, $\xi = \alpha \rho^m$, $m$
being an arbitrary power and $\alpha$ a proportionality constant.

A distinctive feature of the class of solutions  by 
Eq. (\ref{sol1}) is that they describe a violation of the DOC 
which occurs all the time. Namely, we can show  that $ \dot{H}>0$ holds 
over all the times where the solution is defined.

Solutions can be constructed where the violation of the DOC is only 
realized for a finite amount of time. The phase leading to a violation 
of the DOC can then be smoothly connected with an ordinary evolution 
preserving both the DOC and the SEC and 
whose Hubble parameter declines in time.  
In order to have this behavior dynamically realized we have to require that 
the bulk viscosity coefficient will change sign at the moment connected with 
the transition from the phase violating the DOC to the phase where the DOC is
restored. Let us  then assume that $ \xi= q \dot{H}/H$ where $q$ is 
a constant to be determined by consistency with the Friedmann equations. 
The reason for this parameterization is indeed simple. As we stressed, the 
amount of violation of the DOC is indeed proportional to $\dot{H}$. 
Moreover, the  
compatibility with the Eqs. (\ref{fr2}) requires that $q = (2/9) M^2_{P}$.

Under this ansatz, a class of solutions of Eqs. (\ref{fr2}) is given by 
the scale factor
\begin{equation}
a(t) =\exp{[ \delta~ {\rm arcsinh}(t)]}\equiv
 \biggl( t + \sqrt{ t^2 + t^2_1}\biggr)^{\delta},
\end{equation}
whose associated Hubble parameter is given by 
\begin{equation}
H= \frac{\delta}{\sqrt{ t^2 +t^2_1}}.
\end{equation}
We can immediately see by taking the limit of $a(t)$ for
 $t\rightarrow \pm \infty$ we get that 
\begin{equation}
a_{\pm \infty}(t) \rightarrow (\pm t)^{\pm \delta}
\end{equation}
In this model the explicit  evolution of the shear parameter 
and of the energy density can be obtained, according to Eqs. (\ref{fr2}), as 
\begin{equation}
\xi(t)= - \frac{2}{9} M^2_{P} \frac{t}{t^2 + t^2_1},~~~\rho(t)= 
\frac{M^2_{P}~\delta^2}{t^2 + t^2_1}. 
\end{equation}
Notice that $\xi(t)$ 
 is positive for $t\rightarrow -\infty$ and it is negative for 
$t\rightarrow +\infty$. Therefore, the DOC is violated 
in the cosmic time interval  $]-\infty, 0]$
and it is restored for $[0, +\infty[$. The energy density and the curvature
are regular and well defined for every $t$ and, moreover,
$\ddot{a} >0$ for $]-\infty, 0]$ whereas $\ddot{a} < 0$ for $[0, +\infty[$.

If we want the phase where the DOC is violated to be smoothly connected  
to a  radiation dominated evolution  we are led to choose $\delta= 1/2$. With 
this choice we can see from the exact expressions of the scale factor that
\begin{equation}
a(t) = \biggl( t + \sqrt{ t^2 + t^2_1}\biggr)^{\frac{1}{2}},
\label{sfrad}
\end{equation}
Notice that, in this case, the pressure density goes 
as $p \rightarrow \rho/3$ for $t\rightarrow +\infty$.
The examples we just discussed suggest a possible speculation.
The initial phase where the DOC and the 
SEC are both violated could evolve towards a phase where the 
SEC and the DOC are both restored. 

If this is the case we can show that the obtained graviton spectra excited by
 this type of background might be of some phenomenological relevance.
Suppose indeed that the Universe evolves through two  different stages of 
expansion. In the first stage the DOC and the SEC will be both 
violated (i.e. $a(t) \sim (-t)^{-1/2}$) In the second phase 
the DOC and the SEC are both restored (i.e. $a(t) \sim t^{1/2}$). 
If the transition from the regime $p+\rho <0$ to the regime $p+ \rho>0$
occurs smoothly we can also expect the presence of a transition 
epoch where $p + \rho =0 $ and the DOC is saturated but not violated.
Then by ensuring the continuity of the scale 
factors (and of their first derivatives)  between the three  
 regimes we have that the background evolution
 can be written, in conformal time, as 
\begin{eqnarray}
&& a(\eta) = \bigl[-\frac{\eta}{\eta_1}\bigr]^{- \frac{1}{3}},~~~~
\eta \leq -\eta_{1},~~~~~~~\Rightarrow ~p + \rho <0	
\nonumber\\
&& a(\eta) = \bigl[\frac{ 2\eta_1  - \eta}{3 \eta_1}\bigr]^{-1},
~~~~-\eta_{1}< \eta <\eta_2,~~~\Rightarrow ~p + \rho =0
\nonumber\\
&& a(\eta) = \frac{ ( \eta + 2 \eta_1 - \eta_1)\eta_1}{( 2 \eta_1 - \eta_2)^2},
~~~~~~\eta >\eta_2,~~~\Rightarrow ~ p + \rho >0.
\label{scalconf}
\end{eqnarray}

The evolution equation of the proper amplitude of each polarization of 
the tensor fluctuations $h_{ij}$ of a conformally flat background metric of 
Friedmann-Robertson-Walker type can be written, in conformal time and in 
Fourier space, as 
\begin{equation}
\psi'' + \bigl[ k^2  - \frac{a''}{a}\bigr] \psi=0,
\label{grav}
\end{equation}
where for each polarization $\psi = a h$ (the prime denotes derivation 
with respect to conformal time).

The amplification induced by the background 
of Eq. (\ref{scalconf}) in the proper amplitude of the 
gravitational waves  can be obtained from 
the solutions of Eq. (\ref{grav}) in the three temporal 
regions separated by $\eta_1$ and $\eta_2$:
\begin{eqnarray}
&&\psi_{I}( k\eta) = \sqrt{\frac{\pi}{2}} 
 e^{- i \frac{2\pi}{3}} \frac{1}{\sqrt{ 2 k}} 
H^{(2)}_{5/6}(k\eta),~~~~~~~~~\eta\leq -\eta_1,
\nonumber\\
&&\psi_{II}(k\eta) = \sqrt{\frac{\pi}{2}} 
\frac{1}{\sqrt{2 k}}\bigl[e^{-i\pi}
A_{+}(k) H^{(2)}_{3/2}(k\eta) + A_{-}(k) e^{i \pi} 
H^{(1)}_{3/2}(k\eta)\bigr],~~~-\eta_1 <\eta< \eta_2,
\nonumber\\
&&\psi_{III}(k\eta) = \frac{1}{\sqrt{2 k}}\bigl[ B_{+}(k) e^{- i k\eta} 
+ B_{-}(k) e^{ i k \eta}\bigr],~~~~\eta>\eta_2
\label{solpsi}
\end{eqnarray}
where  $H^{(1,2)}_{\nu}$ are the Hankel functions.
By matching in $\eta=-\eta_1$ $\psi_{I,II,III}$ and $\psi'_{I,II,III}$
in $-\eta_1$ and $\eta_2$ we can 
determine the mixing coefficient $A_{\pm}(k)$, $B_{\pm}(k)$ whose frequency 
behavior will fix the spectral evolution of the energy density.

This is a very straightforward
 calculation \cite{10} and the final result can be 
expressed in terms of the energy density of the relic gravitons at the present 
time and in critical units  
\begin{eqnarray}
&&\Omega_{{\rm GW}}(\omega,t_0) = \Omega_{\gamma}(t_0) s^2,~~~~~~~~~~~~
\omega_2  < \omega <\omega_1
\nonumber\\
&&\Omega_{{\rm GW}}(\omega,t_0) =\Omega_{\gamma}(t_0) s^2 
\biggl(\frac{\omega}{\omega_2}\biggr)^{\frac{4}{3}},~~~~~~ 
\omega_{{\rm dec}}<
\omega<\omega_2
\end{eqnarray}
where $s \sim H_1/M_{P}$; $\Omega_{\gamma} \sim 2.6 
\times 10^{-5} h_0^{-2}$ is the present fraction of critical energy density 
stored in radiation and $\omega = k/a = 2 \pi \nu $.

Now our argument is very simple. The only free parameter in this 
model is given by the duration of the intermediate phase saturating the DOC,
or, in other words by the ratio $\omega_1/\omega_2$. The overall 
normalization of the spectrum can be  fixed by assuming, at $\omega_1$, 
the maximal amplitude compatible with the nucleosynthesis bound
$\Omega_{{\rm GW}} \laq 10^{-5}$. Taking into account that 
$\omega_1\sim 10^{11}$ Hz we must conclude that the energy density 
of the relic gravitons at the scale of the interferometric detectors 
(i.e. $\omega_{I} \sim 0.1$ kHz) is given by 
\begin{equation}
\Omega_{{\rm GW}} \sim 10^{-5} 
\bigl(\frac{\omega_{I}}{\omega_{2}}\bigr)^{4/3}.
\end{equation}
Therefore, if the duration of the intermediate phase lies in the range 
$|\eta_2/\eta_1| \sim 10^{7}$--$10^{8}$ we can conclude that 
the energy density of relic gravitons can be as large as 
$10^{-7}$--$10^{-6}$ in the frequency range probed by the interferometers
without conflicting with other bounds coming from lower frequencies.

We showed that the existence of tilted spectra of relic gravitons 
can be connected, in the context of general relativity, with 
the violation of the DOC induced by bulk viscous stresses. In many respects
what we presented in this paper are toy examples which are only
 suggestive of a possible dynamics. In the near future these speculations 
can have some phenomenological impact. Possible indications 
of the existence of tilted gravitons spectra at the frequency of the 
interferometric devices can put bounds on the possible violation of 
the DOC in the early Universe. If tilted gravitons spectra of the type 
discussed in the present paper will be detected, then the conclusion is
twofold. Either the DOC was violated in the early Universe, or the 
theory of gravity underlying the correct model 
describing the early stages of the evolution of the Universe was not of 
Einstein-Hilbert type \cite{11}.

\end{document}